\documentclass[a4paper,preprint]{elsarticle}

\usepackage{epsfig}
\usepackage{graphicx}
\usepackage{amsmath}
\usepackage{rotating}

\begin{document}
\title{Gregarious vs Individualistic Behavior in Vicsek Swarms and the Onset of First-Order Phase Transitions}
\author{Gabriel Baglietto$^{1,2}$, Ezequiel V. Albano$^{1,3}$ 
and Juli\'an Candia$^{1,4}$\\
$^1${\small\it Instituto de F\'{\i}sica de L\'{\i}quidos y Sistemas 
Biol\'ogicos (CONICET, UNLP),}\\{\small\it 59 Nro 789, 1900 La Plata, Argentina}\\
$^2${\small\it Facultad de Ingenier\'{i}a (UNLP), La Plata, Argentina}\\
$^3${\small\it Departamento de F\'{\i}sica, 
Facultad de Ciencias Exactas (UNLP),}\\ 
{\small\it La Plata, Argentina}\\
$^4${\small\it Department of Physics, University of Maryland,}\\ 
{\small\it College Park, MD 20742, USA}}

\begin{abstract}
The Standard Vicsek Model (SVM) is a minimal nonequilibrium model of self-propelled particles that 
appears to capture the essential ingredients of critical flocking phenomena. In the SVM, 
particles tend to align with each other and form ordered flocks of collective motion; however, 
perturbations controlled by a noise term lead to a noise-driven, continuous order-disorder 
phase transition. In this work, we extend the SVM by introducing a parameter $\alpha$ that allows 
particles to be individualistic instead of gregarious, i.e. to choose a direction of motion independently of 
their neighbors. By focusing on the small-noise regime, we show that a relatively small probability of individualistic 
motion (around $10\%$) is sufficient to drive the system from a Vicsek-like ordered phase to a disordered phase. 
Despite the fact that the $\alpha-$extended Model preserves the $O(n)$ symmetry, the interaction range, as well as the dimensionality 
of the underlying SVM, this novel phase transition is found to be discontinuous (first-order),  
an intriguing manifestation of the richness of the nonequilibrium flocking/swarming phenomenon. 
\end{abstract}

\maketitle

\section{Introduction}
Swarming and flocking arise as emergent phenomena of collective motion behavior in a large variety of living and non-living self-propelled particle systems of great interdisciplinary interest. Nature offers a huge variety of    
collective motion phenomena in self-propelled living systems at all scales, from biomolecular micromotors, migrating 
cells and growing bacteria colonies, to insect swarms, fish schools, bird flocks, mammal herds and even human crowds~\cite{vics10}. 
Moreover, many non-living systems of great practical interest involve collective motion and swarming behavior, particularly 
in robotics, where swarms of robots are used in terrain exploration, plague control, optimization of telecommunication networks, 
surveillance and defense, and other tasks without centralized control that appear too challenging to be carried out by an 
individual agent~\cite{vics10}. 
Very recently, the novel concept of chemical robots (also known as {\it chobots}) has been envisioned as an army of millions 
of micrometer-sized robots whose tasks will be to release a chemical payload, or to 
mix two chemical reactants from different compartments within the chobots when they reach their goal~\cite{gran11}. 

Instead of focusing on the 
specific details that make each of these self-propelled systems unique, 
statistical physicists have been studying the general patterns of collective motion, 
aiming to identify the general laws and underlying principles that may govern their behavior.
From this perspective, one important question to address is the onset of ordered macroscopic phases, i.e. the way in which individuals 
having short-range interactions are capable of self-organizing into large-scale cooperative patterns in the absence of leaders or other ordering cues from the environment. By analogy with large molecular systems, flocking and swarming phenomena can be associated 
with phase transitions that depend on a few parameters that characterize the macroscopic states, such as the density of individuals and the flock size. 
For instance, Buhl et al.~\cite{buhl06} investigated the collective motion of locusts, which display a density-driven transition 
from disordered movement of individuals within the group to highly aligned collective motion. Similar transitions have been observed 
in the colective motion of zooplankton swarms~\cite{orde03}, fish schools~\cite{becc06}, and many other self-propelled particle systems 
(\cite{vics10} and references therein).   

On the theoretical side, Vicsek et al.~\cite{vics95} proposed a minimal model to study the onset of order in systems of self-driven individuals, 
which was later followed by other investigations by means of agent-based modeling~\cite{czir99,greg04,alda07}, the Newtonian 
force-equation approach~\cite{mikh99,erdm05}, and the hydrodynamic approximation~\cite{tone95,csah97,tone05}. 
The so-called Standard Vicsek Model (SVM)~\cite{vics95} 
assumes that neighboring individuals tend to align 
their direction of movement when they are placed within a certain interaction range. This alignment rule, which would trivially lead to fully ordered collective motion, is complemented by a second one that introduces noise in the communications among individuals. 
In their seminal paper, Vicsek et al. show that the noise 
amplitude drives the system through a continuous, second-order transition between an ordered phase of collective motion and a disordered phase. 

In this context, the aim of this work is to explore an extension of the SVM, in which an additional parameter 
$\alpha$ is introduced to control the gregarious vs individualistic behavior of particle motion. That is, an individual has a probability $\alpha$ 
to adopt its own direction of motion regardless of its neighbors, and a probability $1-\alpha$ to move according to the SVM's rules. This $\alpha-$extended 
Model may account for ``free will" behavior in biological systems (i.e. the fact that living organisms are not ruled by fixed decision algorithms 
and are therefore able to make unforeseen individualistic decisions at any time) as well as for random failures in robotic and other artificial 
self-propelled particle systems (e.g. the chemically-driven {\it chobots} mentioned above). 
By focusing on the small-noise regime, we show that a relatively small probability of individualistic 
motion (around $10\%$) is sufficient to drive the system from a Vicsek-like ordered phase to a disordered phase. 
Besides the practical interest of extending the well-studied SVM to novel scenarios of particle behavior, we find a theoretically 
intriguing manifestation of the richness of the swarming phenomenon, namely that the $\alpha-$driven 
phase transition is discontinuous (first-order), despite the fact that the $\alpha-$extended Model preserves the $O(n)$ symmetry, 
the interaction range, as well as the dimensionality of the underlying SVM. 

The rest of this paper is organized as follows. In Section 2, we present the definition of the SVM and the $\alpha-$extended Model. 
Section 3 presents a discussion of our results, while our conclusions finally appear in Section 4. 

\section{The Standard Vicsek Model (SVM) and the $\alpha-$extended Model}
 
The SVM~\cite{vics95} consists of a fixed number of interacting particles, $N$, which are moving on a plane. 
In computer simulations, that plane is typically represented by a square of side $L$ with periodic boundary conditions~\cite{vics95,bagl09a}. 
The particles move off-lattice with constant and common speed $v_0\equiv |\vec{v}|$.
Each particle interacts locally adopting the direction of motion of the subsystem of neighboring particles (within an interaction circle of radius $R_0$ centered in the considered particle), 
which is perturbed by the presence of noise. Since the interaction radius is the 
same for all particles, we adopt the interaction radius as the unit of length 
throughout, i.e. $R_0\equiv 1$. The model is implemented as a cellular automaton, 
i.e. all particles update their states simultaneously in one time step.

The updated direction of motion for the $i-$th particle, $\theta_i^{t+1}$, is given by  
\begin{equation}
\theta_i^{t+1}=Arg\left[\sum_{\langle i,j\rangle}e^{i\theta_j^t}\right]+\eta\xi_i^t\ ,
\label{anterm}
\end{equation}
\noindent where $\eta$ is the noise amplitude, the summation is carried over all particles within the 
interaction circle centered at the $i-$th particle, and $\xi_i^t$ is a realization of a $\delta$-correlated white noise uniformly 
distributed in the range between $-\pi$ and $\pi$. 

In this work, we implement the model dynamics by adopting the so-called forward update (FU) rule, 
in which the updated velocities at time $t+1$ determine the new positions according to
\begin{equation}
\vec{x_i}^{t+1}=\vec{x_i}^t+\vec{v_i}^{t+1}\Delta t\ ,
\label{FU}
\end{equation}
where $\Delta t\equiv 1$ is the unitary time step of the cellular automaton. 
This scheme is used for computational convenience, since the transient length is shorter than using the backward update rule that 
determines the new velocities {\it after} the positions are updated. 
In the context of Vicsek model studies, the FU rule was first introduced by Chat\'e et al.~\cite{chat08a,chat08b}. 
As explained in Ref.~\cite{bagl09}, the FU scheme requires to implement an 
algorithm that ensures the $O(n)$ rotational invariance of the SVM, since spurious effects induced by the boundary conditions 
of the simulation space may arise otherwise. At every time step, a frame rotation angle, $\Theta_{fr}$, 
is randomly selected with uniform probability in the $-\pi\leq \Theta_{fr}< \pi$ range. After all particle velocities are accordingly transformed, 
the usual Vicsek model rules are applied. The effects of boundary conditions on the displacement space, as well as alternative methods to 
ensure the $O(n)$ rotational invariance, are further discussed in Refs.~\cite{bagl09,nagy07,alda09}. 

In order to study the relative effects of gregarious vs individualistic behavior, we consider an extension of the SVM model in which 
a parameter $\alpha > 0$ is introduced as follows. When considering the new velocity of a given particle, we first determine whether that particle 
will behave individualistically or gregariously: with probability $\alpha$, the particle will choose a new direction of motion at random (and 
independently of its neighbors), and with probability $1-\alpha$, the particle's new velocity will follow the standard Vicsek's rule, 
Eq.(\ref{anterm}). In the former case, the new direction of motion is randomly chosen with uniform probability in the range between $-\pi$ and $\pi$. 
In passing, let us point out that, in a different context, mixtures of spins coupled to independent heat baths have been studied for the kinetic 
Ising model and far-from-equilibrium spin chains~\cite{garr87,racz94,schm02,hurt04}. 
By analogy, we can consider our model as dealing with off-lattice ferromagnetic ``spins" 
that undergo displacements along the direction of the spin. At any given time, gregarious individuals could be thought of as ``cold spins", while 
individualistic individuals could be analogously regarded as ``hot spins" coupled to a thermal bath at infinite temperature.  
Notice also that, although more complex variations of the SVM could be considered, we purposefully focus on the simplest 
extension of the SVM that includes a ``free-will" option (or ``failure" option, if considered within the context of non-living systems). 
Indeed, it is our aim to preserve the minimal nature of the SVM and show that, from the point of view of nonequilibrium statistical mechanics, 
a qualitatively novel and distinct behavior arises from this simple extension. Moreover, the SVM is fully recovered by simply letting $\alpha\to 0$. 

The parameters of the model are the number of particles $N$, the dimension of the displacement space $d$, the linear size of the displacement space $L$, the interaction radius $R_0$, 
the particle density $\rho$, the particle speed $v_0$, the noise amplitude $\eta$, and the individualistic probability $\alpha$. 
Notice, however, that $\rho=N/L^d$, 
where $d=2$ in the standard case, and that $R_0\equiv 1$ can be chosen as the unit of length. Moreover, $v_0^{-1}$ merely plays 
the role of a ``thermalization parameter" that 
measures how many times, on average, two neighbors check out each other's positions while they remain at a distance within the unitary interaction radius. Therefore, 
relative large values of speed (typically $v_0\geq 0.3$) correspond to the low-thermalization regime, characterized by highly anisotropic diffusion and the manifestation of simulation artifacts in the form of directionally quantized density waves (see Ref.~\cite{nagy07} for details). In order to avoid spurious artifacts, we choose $v_0=0.1$ and the density value $\rho=0.25$ throughout, which are standard values in the Vicsek 
model literature~\cite{vics95}. 
Moreover, in order to pinpoint effects due to the $\alpha-$extension, we consider the small noise regime ($\eta\ll 1$) throughout 
and focus on the behavior of 
the system over a wide range of values of the parameter $\alpha$. Also, finite-size effects are assessed by studying 
several system sizes in the range $16384\leq N\leq 131072$. 

In our simulations, 
one time step is defined by the simultaneous update of velocities and positions for all $N$ particles in the system. 
For each set of parameter values, time series of $7\times 10^6$ time steps have been generated. 
All observables measured here are statistical averages within the steady-state regime, which is typically attained after $1-2\times 
10^6$ time units. Let us also point out that, for the largest 
system sizes studied here (i.e. $N>10^5$), 
generating this large number of steady-state configurations requires a significant computational effort. 

\section{Results and Discussion}   

The well-studied SVM exhibits a far-from-equilibrium continuous phase transition between ordered states of motion at 
low noise levels and disordered motion at high noise levels. The 
natural order parameter is given by the absolute value of the 
normalized mean velocity of the system~\cite{vics95}:
\begin{equation}
\varphi = \frac{1}{N v_0}\vert\sum_{i=1}^{N}\vec{v_i}\vert,
\label{orpa}
\end{equation}
\noindent where $\varphi$ is close to zero in the disordered phase and close to unity in the ordered phase. 
In order to analyze the critical nature of the SVM, this noise-driven phase transition has been investigated  
by several approaches, e.g. finite-size scaling analysis and short-time critical dynamic techniques~\cite{bagl08,bagl12}. 

\begin{figure}[t!]
\centerline{\epsfysize=3.3in\epsfbox{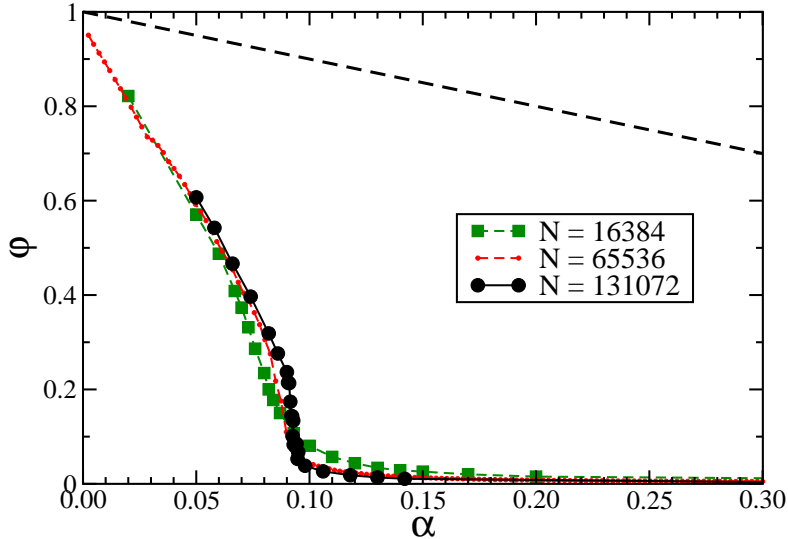}}
\caption{Plots of the order parameter $\varphi$ versus the parameter $\alpha$ (which measures the 
probability for individualistic behavior) for different system sizes, as indicated. By varying $\alpha$, 
the system is driven across an order-disorder transition with very mild finite-size effects. 
The dashed straight line shows the mean-field behavior for $\eta=0$ and $\rho\to\infty$ in the thermodynamic limit, 
where the effective interaction becomes long-range and the order-disorder transition is 
exactly located at $\alpha_t=1$.}
\label{order_par}
\end{figure}

In this work, we explore the $\alpha-$extended Vicsek model, where the parameter $\alpha$ measures the probability 
for individualistic behavior (see Section 2 above for details). Since we focus on the small-noise regime, 
the $\alpha\to 0$ limit corresponds to the SVM ordered phase. 
Figure~\ref{order_par} shows the order parameter $\varphi$ as a function of the parameter $\alpha$. 
We observe that a relatively small probability of individualistic 
motion, $\alpha_t\approx 0.09$, is sufficient to drive the system from a Vicsek-like ordered phase ($\varphi > 0$) 
to a disordered phase ($\varphi\approx 0$). 
Moreover, finite-size effects are mild, as indicated by the near-overlap of the plots of different system sizes in the range 
$16384\leq N\leq 131072$. The absence of significant finite-size effects near the transition region at 
$\alpha_t\approx 0.09$ suggests that the phase transition may be discontinuous (first-order) instead of the continuous 
(second-order) phase transition characteristic of the SVM. Indeed, the Binder cumulant analysis 
presented below will provide very strong evidence to confirm the first-order nature of the 
phase transition and will also allow us to determine the transition value with great accuracy.   
The dashed straight line 
in Figure~\ref{order_par} shows the mean-field limit of $\varphi$ vs $\alpha$ for $\eta=0$ when all particles interact with each other, 
i.e. when $\rho\to\infty$ in the thermodynamic limit ($N\to\infty$). In this limiting case, a fraction $\alpha$ of the system is formed by 
particles moving in random directions and therefore has a null contribution to the order parameter, while a fraction $1-\alpha$ is fully 
aligned along the mean direction of motion of the whole system. Thus, $\varphi=1-\alpha$. It is interesting to note that, in this     
limit of infinite density where the effective interaction becomes long-range, the transition from order to disorder is 
exactly located at $\alpha_t=1$.

\begin{figure}[t!]
\centerline{\epsfysize=1.8in\epsfbox{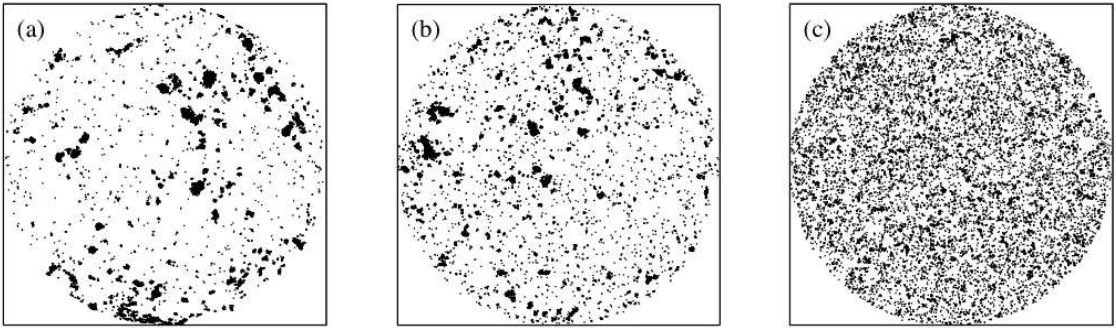}}
\caption{Typical snapshot configurations for $N=16384$ and different values for 
$\alpha$, namely: (a) $\alpha=0.02$, (b) $\alpha=0.093$, and (c) $\alpha=0.20$. These configurations have been generated 
by applying the algorithm proposed in Ref.~\cite{bagl09}, in which the $O(n)$ rotational invariance is enforced explicitly 
in the model's dynamic rules.}
\label{snapshots}
\end{figure}

Further qualitative insight can be gained by looking at the clustering of particles in 
typical snapshot configurations across the transition. Notice that the standard order parameter for Vicsek-like flocking phenomena, 
Eq.(\ref{orpa}), refers to the overall direction of motion of the particles. Therefore, the spatial distribution of particles offers a 
related, complementary view of the degree of order in the system.  
Figure~\ref{snapshots} shows snapshot configurations for a system of size $N=16384$ and different values for the 
individualistic probability $\alpha$, namely: 
(a) $\alpha=0.02$, (b) $\alpha=0.093$, and (c) $\alpha=0.20$. 
The low-$\alpha$ regime (Fig.~\ref{snapshots}(a)) 
is characterized by the existence of large flocks  
that are responsible for a system-wide, large-scale coherent motion. Inbetween flocks, some regions display 
relatively large voids in which the particle density is very low. At the transition (Fig.~\ref{snapshots}(b)), 
the ordered flocks appear scattered 
throughout the system in coexistence with the disordered phase. Notice also that flock sizes do not appear to depart 
significantly from the average flock size, i.e. the snapshot suggests that the flock size distribution does not follow a power law 
(as would be expected if the transition were continuous). Finally, the high-$\alpha$ regime (Fig.~\ref{snapshots}(c)) shows 
a configuration where individuals are nearly evenly distributed across the system and large flocks are absent, 
as expected for the disordered phase. Thus, at a qualitative level, we observe that the spatial distribution of individuals 
exhibits different configurations that, as the parameter $\alpha$ is increased, change from ordered to disordered through a 
state of phase coexistence. Quantitative evidence for this first-order phase transition is provided by the so-called Binder 
cumulant of Vicsek's order parameter, as follows. 

The Binder cumulant, defined by
\begin{eqnarray}
U_4=1-\frac{\langle \varphi^4\rangle}{3\langle \varphi^2\rangle^2}\  ,
\label{binder_eq}
\end{eqnarray}
\noindent is a fourth-order cumulant dependent on the variance and the kurtosis of the order parameter probability 
distribution. Since, for second-order phase transitions, the scaling prefactor of the cumulant is independent of the 
sample size, plots of $U_4$ versus the control parameter 
lead to a common (size-independent) intersection point that corresponds to the location of the critical 
value of the order parameter in the thermodynamic limit~\cite{bind81}. Moreover, for first-order phase transitions, 
a characteristic signature is for $U_4$ to have a sharp fall towards negative values~\cite{bind84,chal86}. 
This hallmark behavior is somewhat elusive 
because it only appears for relatively large system sizes.   

\begin{figure}[t!]
\centerline{\epsfysize=3.3in\epsfbox{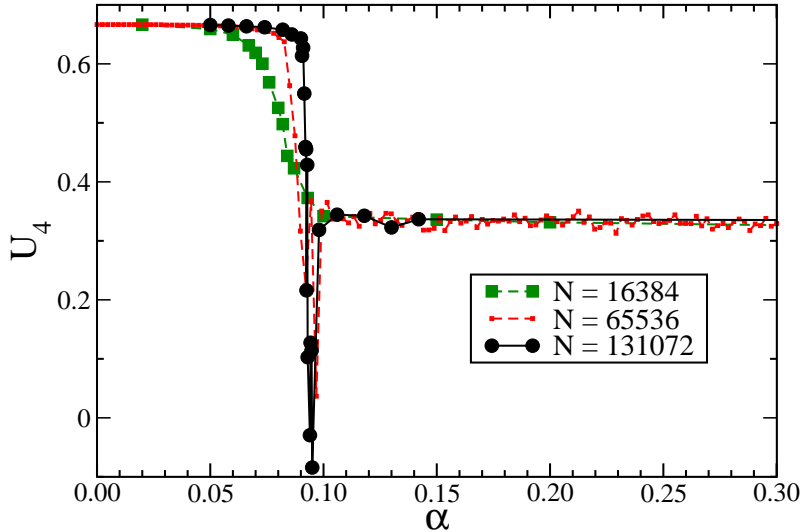}}
\caption{Binder fourth-order cumulant $U_4$ as a function of the individualistic probability $\alpha$ for different system sizes, as indicated.}
\label{binder}
\end{figure}

Figure~\ref{binder} displays 
the Binder cumulant as a function of the individualistic probability $\alpha$ for different system sizes, as indicated. 
In agreement with previous figures, the cumulant's behavior shows the existence of an $\alpha-$driven phase 
transition. Furthermore, although not apparent in the small system size ($N=16384$), the results for 
the larger systems studied here ($N\ge 65536$) show sharp minima at the transition point, which, as discussed above, is the 
signature behavior of discontinuous (first-order) phase transitions. The sharpness of these minima allows us to determine accurately 
the location of the transition at $\alpha_t= 0.0925\pm 0.010$. 

\begin{figure}[t!]
\centerline{\epsfysize=3.3in\epsfbox{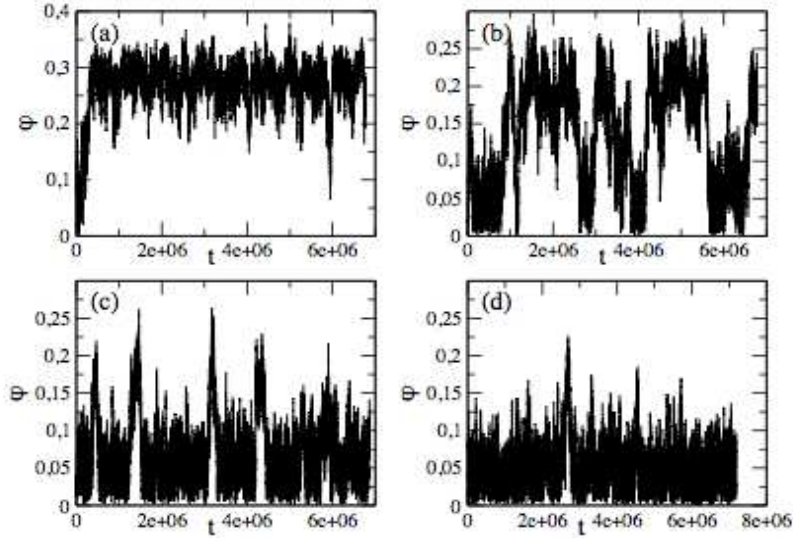}}
\caption{Time series of the order parameter $\varphi$ for a system of size $N=131072$ and different values of $\alpha$ across the transition region.   
(a) For $\alpha=0.086$, the system is ordered and the order parameter displays small fluctuations around $\varphi\approx 0.25-0.30$. 
(b) For $\alpha=0.0928$, strong excursions are observed between one ordered (meta)stable state at $\varphi\approx 0.20$ and one 
disordered (meta)stable state at $\varphi\approx 0.05$. (c) For $\alpha=0.093$, the ordered metastable state is less likely than the 
disordered one. (d) For $\alpha=0.0947$, the system stays chiefly in the disordered state; excursions to the ordered metastable state 
become shorter in duration and less frequent.}
\label{time_series}
\end{figure}

A hallmark behavior of a discontinuous phase transition is the phenomenon of metastability. In equilibrium systems, this phenomenon results from 
the coexistence of two or more local minima of the free energy, where each minimum corresponds to one of the coexisting phases at the transition. 
In finite-size systems, fluctuations may drive the system from one local minimum to another one across the free energy barrier separating those 
(meta)stable states. In our nonequilibrium model, an analogous behavior is observed in time series of the order parameter. Figure~\ref{time_series} 
shows the order parameter $\varphi$ for a system of size $N=131072$ as a function of time for different values of $\alpha$ across the transition region. 
Fig.~\ref{time_series}(a) corresponds to $\alpha=0.086$, i.e. below the phase transition. The system is pinned at the ordered phase and displays 
relatively small fluctuations around $\varphi\approx 0.25-0.30$. Notice also the transient regime from an initially fully disordered state a $t=0$ to 
the steady-state regime, which in this case is reached after $t\approx 5\times 10^5$. Fig.~\ref{time_series}(b) corresponds to $\alpha=0.0928$, where 
strong excursions are observed between one ordered (meta)stable state at $\varphi\approx 0.20$ and one 
disordered (meta)stable state at $\varphi\approx 0.05$. By increasing slightly the control parameter to $\alpha=0.093$, Fig.~\ref{time_series}(c) shows that 
the ordered metastable state becomes less likely than the disordered one. This strong sensitivity to small variations in $\alpha$ is 
in agreement with the sharp transition displayed by the Binder cumulant (Fig.~\ref{binder} above). 
By increasing the control parameter further to $\alpha=0.0947$, as shown by Fig.~\ref{time_series}(d), the 
system stays mostly in the disordered state. In fact, excursions to the ordered metastable state become noticeably shorter in duration and much more 
sporadic. 

\begin{figure}[t!]
\centerline{\epsfysize=3.3in\epsfbox{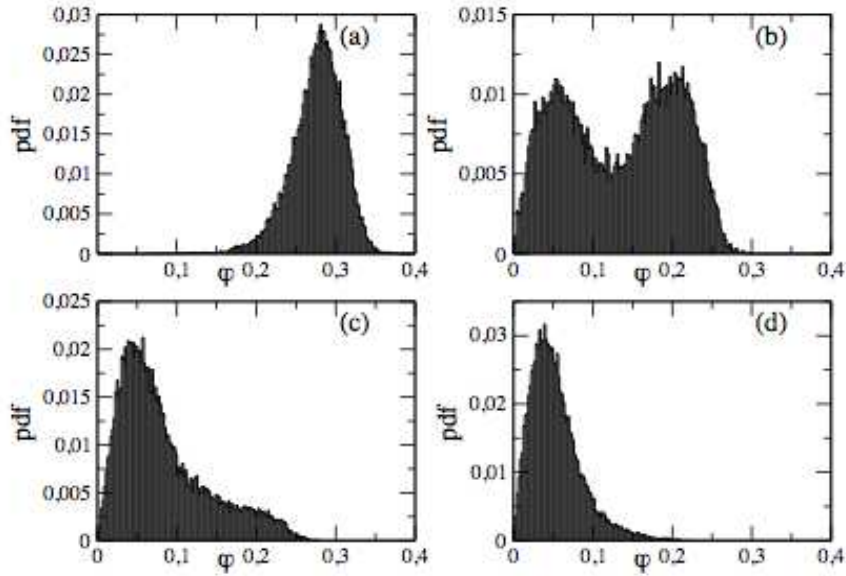}}
\caption{Probability distribution functions (pdf) of the order parameter for a system of size $N=131072$ and different values of 
$\alpha$ across the transition region.   
(a) For $\alpha=0.086$, the system is ordered and the pdf is unimodal and peaked around $\varphi\approx 0.25-0.30$. 
(b) For $\alpha=0.0928$, the distribution is bimodal and corresponds to the coexistence of two distinct phases.
(c) For $\alpha=0.093$, the peak corresponding to the disordered phase prevails, 
although the distribution is still bimodal.
(d) For $\alpha=0.0947$, the peak associated with the ordered phase has vanished and the distribution is completely dominated by 
the peak associated with the disordered phase.} 
\label{OP_pdf}
\end{figure}

The phenomenon of metastability is also reflected in the behavior of the 
probability distribution functions (pdf) of the order parameter across the transition region.        
Figure~\ref{OP_pdf} shows the order parameter probability distribution functions for a system of size $N=131072$ and the same values of $\alpha$ used 
in the previous Figure. In Fig.~\ref{OP_pdf}(a), the distribution for $\alpha=0.086$ 
is unimodal and centered around $\varphi\approx 0.25-0.30$, i.e. the order parameter 
has small fluctuations around the positive mean value that characterizes the ordered phase. Fig.~\ref{OP_pdf}(b), which corresponds to the transition value 
$\alpha=0.0928$, is a bimodal distribution with two similar peaks associated with the coexistence of the ordered and disordered phases. 
In Fig.~\ref{OP_pdf}(c), which was obtained with $\alpha=0.093$, one observes that the peak corresponding to the disordered phase becomes dominant, 
although the distribution is still bimodal. Finally, Fig.~\ref{OP_pdf}(d) represents the disordered regime for $\alpha=0.0947$, where the local maximum 
associated with the ordered phase has completely vanished. As a summary, the panels of Figure~\ref{OP_pdf} display the occurrence of a sharp 
order-disorder transition that takes place within a very narrow range of values of $\alpha$. The hallmark of a discontinuous phase 
transition, in which the two phases coexist, becomes evident by the bimodal probability distribution function of the order parameter 
at the transition, i.e. $\alpha\approx \alpha_t$. Below the transition within the ordered phase $(\alpha<\alpha_t)$, the peak corresponding to the disordered phase vanishes; 
and viceversa, above the transition within the disordered phase $(\alpha>\alpha_t)$, the peak associated with the ordered phase vanishes and the peak corresponding to 
the disordered phase prevails.

\begin{figure}[t!]
\centerline{\epsfysize=3.3in\epsfbox{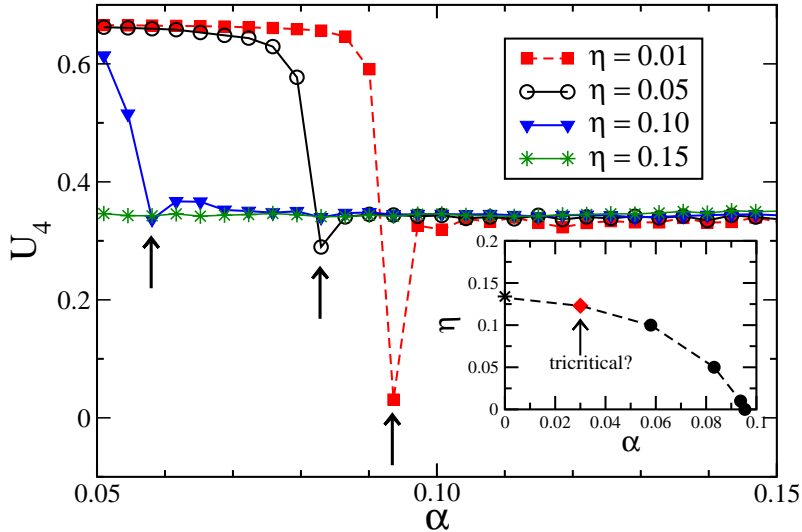}}
\caption{Binder fourth-order cumulant $U_4$ as a function of the individualistic probability $\alpha$ for $N=131072$ and different noise amplitudes, as indicated. Arrows mark the occurrence of minima. Inset: phase diagram in the $\eta$ vs $\alpha$ parameter space, where 
filled circles indicate first-order transitions, the star shows the pure-Vicsek limit with a second-order transition, and the red 
diamond shows a hypotetical tricritical point. See more details in the text.}
\label{binder_eta}
\end{figure}

So far, we have focused on the effects of adding individualistic motion, as controlled by the parameter $\alpha$, on Vicsek-like self-propelled 
particles within the small-noise regime ($\eta\ll 1$). Let us now investigate the behavior of the $\alpha-$extended Model under non-negligible 
noise amplitudes. Figure~\ref{binder_eta} shows the Binder cumulant as a function of the individualistic probability $\alpha$ for $N=131072$ and different noise amplitudes, as indicated. For very small noise amplitudes, e.g. $\eta=0.01$, the Binder cumulant shows a sharp minimum 
at the transition, which as we have discussed above, is a characteristic signature of first-order phase transitions (recall Fig.~\ref{binder}). As the noise amplitude is increased, however, the minima become less noticeable and appear shifted towards smaller values 
of $\alpha$. Increasing the noise amplitude beyond the pure-Vicsek critical transition, i.e. $\eta>\eta_c=0.134$~\cite{bagl08}, 
all signatures of a transition vanish and the Binder cumulant appears flat, no longer depending on the parameter 
$\alpha$. Based on these findings, we are able to sketch a phase diagram in the $\eta$ vs $\alpha$ parameter space, as displayed in the 
Inset to Fig.~\ref{binder_eta}. On the one hand, transition points with characteristic first-order signatures are shown by filled circles. On the other hand, the collective motion transition in the pure-Vicsek limit with $\alpha=0$ has been characterized as second-order~\cite{vics95}; for the velocity and density parameters used in this work, the continuous second-order 
phase transition has been located at $\eta_c=0.134$~\cite{bagl08} (shown by a star in the Inset to Fig.~\ref{binder_eta}). 

Certainly, the fact that the smooth $\alpha-$extension of the SVM changes the critical nature of the system is very intriguing. 
Indeed, these findings are a display of the richness of the swarming phenomenon even when represented by means of a minimal 
nonequilibrium model such as Vicsek's. 
As an example of a rich equilibrium minimal model, in which the critical nature of the underlying system is modified by the 
addition of a Hamiltonian term that preserves the model symmetries, the interaction range and the dimensionality, let us 
mention the Blume-Capel extension of the classical Ising model~\cite{blum66,cape66}. The Blume-Capel model adds a symmetry-preserving term 
proportional to a crystal field $D$ that favors impurities ($S=0$ states) over up/down spins ($S=\pm 1$ states). In the $D\to -\infty$ 
limit, the Blume-Capel model tends smoothly to the classical two-state Ising model. However, in some regions of the $T-D$ parameter space, 
the order-disorder phase transition becomes discontinuous (first-order) and a tricritical point can be determined. 
Reasoning by analogy, and based on the results discussed above, we therefore conjecture the existence of a tricritical point (as indicated 
in the Inset to Fig.~\ref{binder_eta}), which would separate the standard noise-driven continuous transition from the $\alpha-$driven discontinuous transition reported here. The existence of such phenomenon, however, remains an open question that lies 
beyond the scope of the present work.     

\section{Conclusions}
In this work, we investigated the so-called $\alpha-$extended Vicsek Model, in which a parameter 
$\alpha$ was introduced to control the gregarious {\it vs} individualistic behavior of particle motion. 
By focusing on the small-noise regime, we showed that a relatively small probability of independent particle 
motion was sufficient to drive the system from a Vicsek-like ordered phase to a disordered phase. Indeed, the order parameter 
and its associated 
fourth-order Binder cumulant showed very strong evidence of a discontinuous, first-order phase transition occurring at $\alpha_t =0.0925\pm 0.010$. 
Further qualitative and quantitative insight into the nature of the transition was provided by snapshot configurations, time series 
of the order parameter, as well as probability distribution functions of the order parameter. 

The aims of the present study were two-fold. On the one hand, by extending the well-studied Standard Vicsek Model (SVM) to novel scenarios of particle behavior 
(while preserving, at the same time, its nature as a minimal model), we may account for important features of real self-propelled particle systems 
that are not captured by the SVM, as for instance free-will behavior in biological systems and random failures in robotic and other artificial systems. 
Moreover, on the other hand, we found a theoretically 
intriguing manifestation of the richness of the swarming phenomenon, namely that the $\alpha-$driven 
phase transition was discontinuous (first-order), despite the fact that the $\alpha-$extended Model preserved the $O(n)$ symmetry, 
the interaction range, as well as the dimensionality of the underlying SVM. 

By analogy to equilibrium systems such as the Blume-Capel extension of the classical Ising model, 
we conjecture the existence of a tricritical point in the 
$\eta-\alpha$ parameter space of the $\alpha-$extended Model, which would separate the standard noise-driven continuous 
transition from the $\alpha-$driven discontinuous transition reported here. Indeed, the existence of such phenomenon remains a 
very interesting open question that certainly deserves further investigation.

\section*{Acknowledgments}
This work was financially supported by  CONICET, UNLP and  ANPCyT (Argentina).

\end{document}